\documentclass[aps,prl,reprint,amsmath,groupedaddress,twocolumn]{revtex4-1}
\usepackage[pdftex]{graphicx}
\usepackage{graphicx,afterpage}
\usepackage{amsfonts}
\usepackage{amssymb}
\usepackage{graphicx}

\begin{document}


\title{Low-Pump-Power, Low-Phase-Noise, and Microwave \\ 
to Millimeter-wave Repetition Rate Operation in Microcombs}


\author{Jiang Li, Hansuek Lee, Tong Chen and Kerry J. Vahala}
\email[]{vahala@caltech.edu}
\affiliation{T. J. Watson Laboratory of Applied Physics, California Institute of Technology, Pasadena, California 91125, USA}


\date{\today}

\begin{abstract}
Microresonator-based frequency combs (microcombs or Kerr-combs) can potentially miniaturize the numerous applications of conventional frequency combs. A priority is the realization of broad-band (ideally octave spanning) spectra at detectable repetition rates for comb self referencing. However, access to these rates involves pumping larger mode volumes and hence higher threshold powers. Moreover, threshold power sets both the scale for power per comb tooth and also the optical pump. Along these lines,  it is shown  that a class of resonators  having surface-loss-limited Q factors can operate over a wide range of repetition rates with minimal variation in threshold power.  A new, surface-loss-limited resonator illustrates the idea. Comb  generation on mode spacings  ranging from 2.6 GHz to 220 GHz with overall low threshold power (as low as 1 mW) is demonstrated.   A record number of comb lines for a microcomb (around 1900) is also observed with pump power of 200 mW. The ability to engineer  a wide range of repetition rates with these devices is also used to investigate a recently observed mechanism in microcombs associated with dispersion of subcomb offset frequencies. We observe high-coherence, phase-locking in cases where these offset frequencies are small enough so as to be tuned into coincidence.  In these cases, a record-low microcomb phase noise is  reported at a level comparable to an open-loop, high-performance microwave oscillator.
\end{abstract}


\maketitle

The combination of high optical Q factor and the Kerr effect optical nonlinearity enables optical parametric oscillation in microcavities \cite{vahalaOPO, Maleki1}.  Concomitant, non-degenerate four-wave-mixing multiplies the initial Stokes and anti-Stokes waves creating a  route to generation of frequency combs \cite{TJK1}.  This approach offers the potential for miniaturization and integration with other devices, thereby magnifying the already remarkable impact of frequency combs on science and metrology \cite{TJKreview, Newbury}. So far, microcombs  (or Kerr combs) have been demonstrated using silica micro-toroids \cite{TJK1}, CaF$_2$ diamond-milled rods \cite{Maleki2,Maleki3},  fiber Fabry-Perots \cite{Scott1}, silicon-nitride rings on silicon \cite{Gaeta1, Purdue}, high-index silica rings on silicon \cite{Moss} and fused-quartz cavities \cite{Papp}. Octave span operation has been demonstrated in microtoroids \cite{TJK octave}  and in silicon nitride resonators \cite{Gaeta2}, with line spacings of 850 GHz and 226 GHz, respectively; and microwave-repetition-rate is possible in a range of devices \cite{TJK 86GHz, Maleki2, Maleki3, Papp, OPO FIO, Gaeta3}. However, the combination of these properties, required for self-referenced operation, has not been possible.  Moreover,  microwave-rate devices are also prone to operate in a mode whereby oscillation occurs first on non-native comb line separations \cite{Herr}.  This creates a situation in which many subcombs can ultimately oscillate on the native comb spacing, but, significantly, not necessarily with the same underlying offset frequencies.  This dispersion in offset frequencies is now believed to contribute to instability in the microwave beat note \cite{Herr}.

\begin{figure*}[!htbp]
\includegraphics[width=\textwidth]{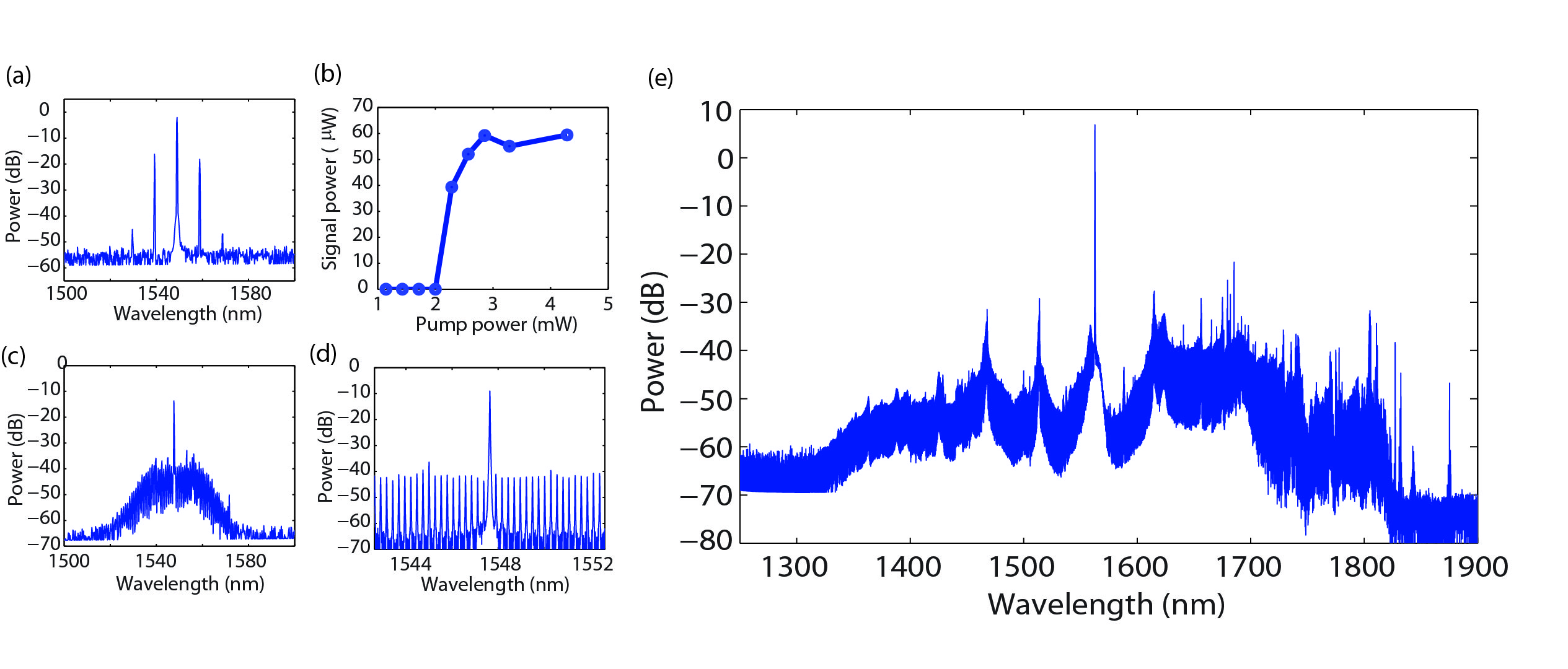}
\caption{\label{fig:comb} Panels (a)-(e) show data obtained using a 2 mm diameter (33 GHz FSR) disk microcomb device. (a) The 2 mm disk comb spectrum is measured for excitation just above threshold. (b) The power of the first oscillating, higher-frequency, comb line is plotted versus pump power  and shows a threshold turn-on power of approximately 2 mW.  (c) Approximately 200 comb lines are generated with coupled pump power of 7.5 mW. (d) A reduced span scan of the spectrum in (c) is shown with comb lines resolved by the OSA. (e) A broadband comb spectrum with 62 THz span is shown. The coupled pump power was 200 mW. }
\end{figure*} 

In this work, both of these problems related to microwave-rate systems are investigated using a new optical resonator. It  is silica-based on a silicon chip and provides Q factors as high as 875 million \cite{DiskNatureP}. Other properties of the device, including surface roughness measurements, are given in \cite{DiskNatureP}. Because the devices are lithographically defined and achieve ultra-high-Q operation without the need for a reflow process \cite{toroid}, microcomb operation is achieved across a record span (2.6 GHz - 220 GHz) of user-defined, repetition rates. Indeed, the rates presented here are the lowest achieved to date for any microcomb. Moreover, the devices are surface-loss-limited over a wide range of diameters, a property that is  shown to approximately decouple a strong dependence of pumping threshold on repetition rate so that turn-on power remains less than 5 mW for repetition rates between 4.4 GHz and 220 GHz.

In the experiment, an external-cavity diode laser in the C band is amplified using an erbium-doped fiber amplifier (EDFA) and then coupled to the disk resonator using a tapered fiber coupler \cite{taperfiber1,taperfiber2}. The pump (TE polarized) is thermally-locked to the resonance \cite{thermallock}.  The generated comb lines from the disk resonator are then coupled to the same taper fiber through which spectral monitoring or photo-detection (demodulation) is straightforward. Spectral monitoring is performed using both a telecom optical spectrum analyzer (OSA) (600 nm - 1700nm) and an infrared OSA (1200 nm - 2400 nm). Comb lines are demodulated on a high-speed photodetector having a bandwidth of 25 GHz. The resulting photocurrent beat notes are analyzed on an electrical spectrum analyzer  and also using a phase noise analyzer. 

\begin{figure}[!b]
\includegraphics[width=0.46\textwidth]{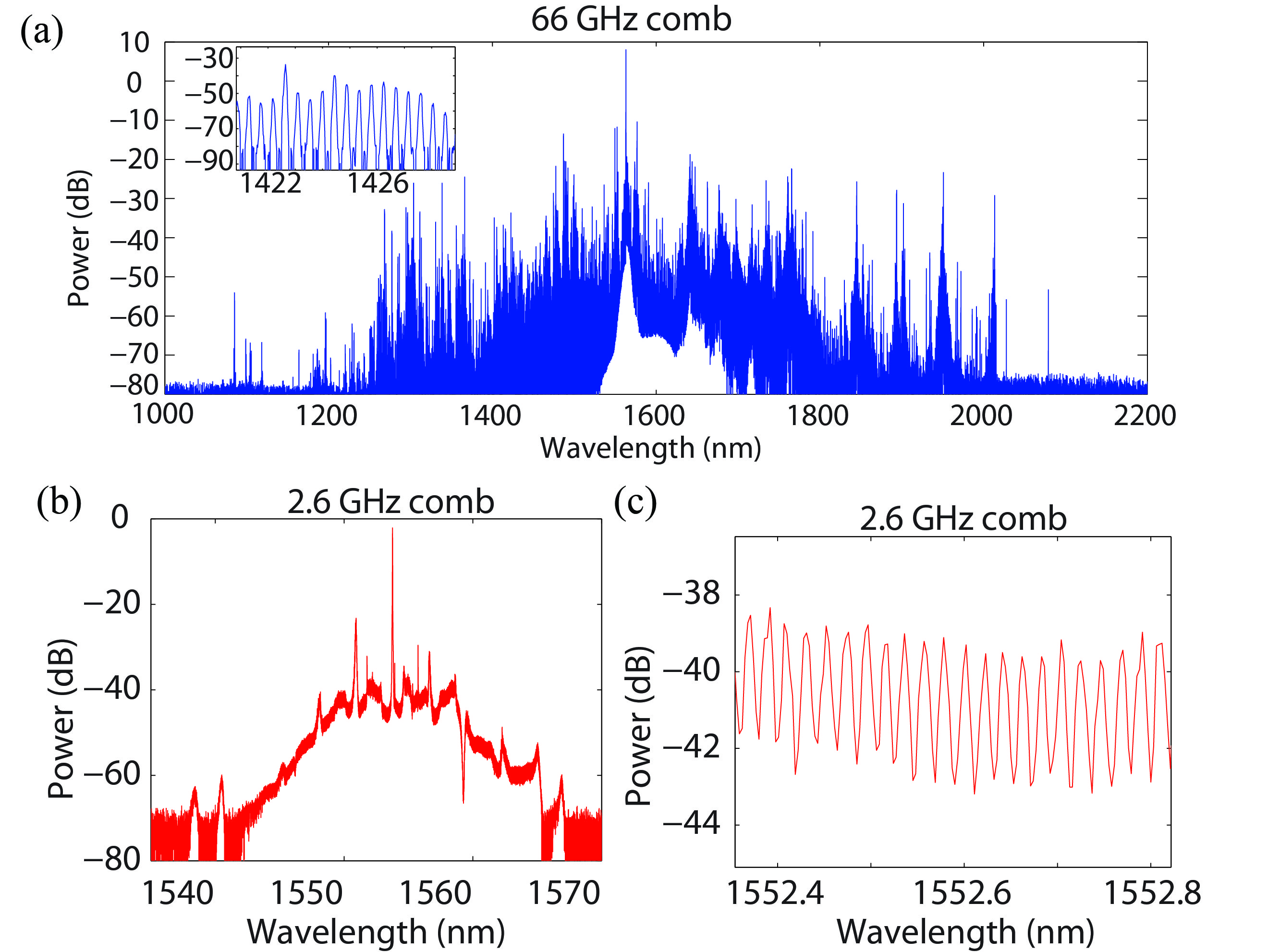}
\caption{\label{fig:reprate} (a) A broadband comb spectrum with 106 THz span (1180 nm - 2020 nm) and repetition rate 66 GHz is shown. Inset: Reduced-span spectrum of the comb with comb lines resolved. (b) An optical comb spectrum taken using a  2.6 GHz FSR device. (c) A reduced wavelength span of the spectrum in (b)  in which the individual comb lines are just resolved by the OSA. }
\end{figure}

In Fig.~\ref{fig:comb}(a), the spectrum of a 2 mm diameter disk (free-spectral-range, FSR, is 33 GHz) having a threshold of 2 mW (Fig.~\ref{fig:comb}(b)) is shown for excitation only slightly above threshold. This resonator featured an intrinsic Q of 270 million. The initial Stokes and anti-Stokes lines occur at a multiple of 37 of the native line spacing, which is consistent with estimate of the parametric gain spectral maximum based on calculated dispersion and known cavity Q (see discussion in \cite{Herr}). Only modest pumping above threshold is required to generate a dense comb spectrum on the native line spacing. About 200 comb lines are generated with coupled pump power of only 7.5 mW (Fig.~\ref{fig:comb}(c) and ~\ref{fig:comb}(d)). Further increase of the coupled pump power to 200 mW leads to a broad-band comb spectrum from 1320 nm to 1820 nm (Fig.~\ref{fig:comb}(e)). This spectrum spans nearly half an octave (62 THz) and contains  about 1900 comb lines, which is to the author¡¯s knowledge the largest number of comb lines so far generated from a microcomb. Using a larger FSR device (66 GHz), it was possible to obtain 3/4 octave span operation in a continuous spectrum (106 THz, 1180 nm - 2020 nm, see Fig.~\ref{fig:reprate} (a)).  Spectra taken for a microcomb featuring a record-low FSR of 2.6 GHz are also shown in Fig.~\ref{fig:reprate}(b) and \ref{fig:reprate}(c).

The threshold relation for parametric oscillation in a microcavity \cite{vahalaOPO} can be manipulated into the following form,
\begin{equation} \label{eq:thres}
P_{th}	\approx \frac{\pi}{8\eta}\frac{n}{n_2}\frac{\omega}{\Delta\omega_{FSR}}\frac{A}{Q_T^2}
\end{equation}                                                       
{\noindent}where $\eta = \kappa_e/\kappa$ is the coupling parameter ($\kappa$ and $\kappa_e$ are the total and coupling-related cavity decay rates), $n_2$ ($n$) is the nonlinear index (refractive index), $\Delta\omega_{FSR}$ ($\omega$) is the  free-spectral-range (optical frequency), $A$ is the mode area, and $Q_T$ is the total optical Q factor. In the present device, $n_2 = 2.2\times10^{-20} $m$^2$/W (silica) and $A\sim 30 \mu$m$^2$ for a 2 mm disk cavity. All other factors held fixed, it is clear that decreasing FSR (to achieve microwave-rate comb operation) adversely impacts turn-on power. Moreover, in whispering-gallery resonators, the mode area, $A$, will generally increase with decreasing FSR, thereby causing further degradation of power requirements. At the same time, it is interesting to note the positive impact of increasing Q factor.  Higher optical Q creates larger resonant build-up so that a given coupled power induces a larger Kerr nonlinear coupling of signal and idler waves. It also reduces oscillation threshold since optical loss is  lowered.  The combined effect leads to the inverse quadratic behavior in Eq.(\ref{eq:thres}). 

The dependence of optical Q in whispering-gallery resonators versus resonator diameter (FSR) depends strongly on whether the round-trip losses are set by surface or material losses. In the material-loss limit, Q is independent of  FSR and turn-on power will therefore degrade as a  comb transitions from millimeter-wave to microwave repetition rates. However, in surface-loss-limited resonators, the resonator Q factor increases (nearly proportionally) with resonator diameter. This happens because the optical field strength at the resonator surface decreases with increasing resonator diameter. In such cases, the surface-loss-limited behavior of Q offsets the FSR and area dependences, leading to low, turn-on power (and total required power) across a wide range of repetition rates. 

\begin{figure}[t]
\begin{center}
\includegraphics[width=0.45\textwidth]{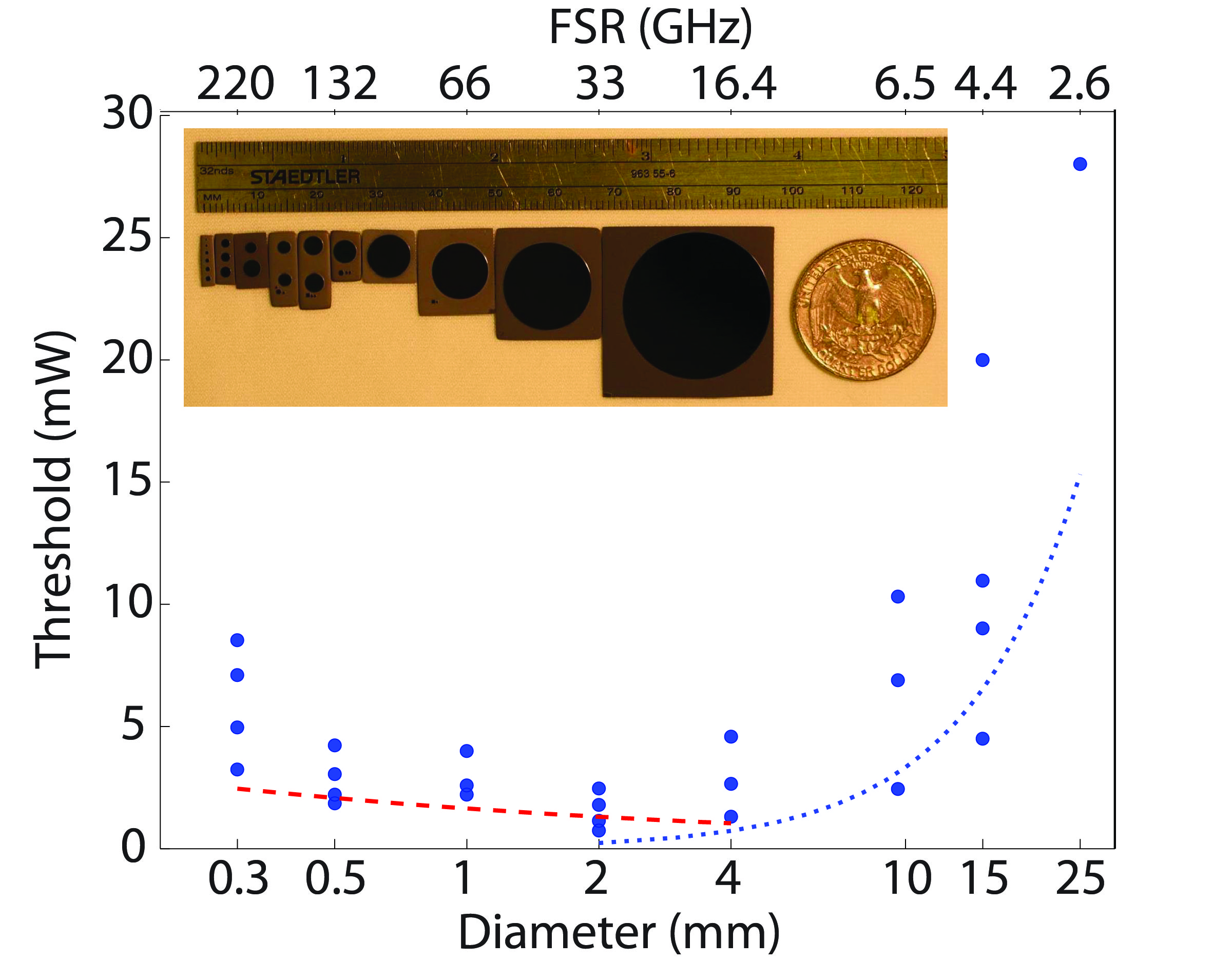}
\caption{\label{fig:schematic} Plot showing measured microcomb threshold versus resonator diameter (lower axis) and FSR (upper axis).  Also shown are the trend curves of the threshold data using $D^{-1/3}$ (dashed curve) and  $D^{5/3}$ (dotted curve). Inset:  A photograph of microcomb disk resonators ranging in diameter  from 0.3 mm to 25 mm, all of which provide parametric oscillation and comb generation. }
\end{center}
\end{figure}

To study the effect of surface-loss-limited Q behavior on threshold, a series of devices having a range of diameters were fabricated and characterized.
Fig.~\ref{fig:schematic} shows a plot of the parametric oscillation threshold versus diameter (lower axis) and FSR (upper axis) for several of the devices. Threshold levels associated with  different resonator pump modes  are displayed and measured variations result from  changes in both Q factor and modal area for each pump mode. The lower-order transverse modes of the device are expected to give the lower threshold turn-on power since these have  higher Q and smaller mode area \cite{DiskNatureP}. Microcomb oscillation at rates from 2.6 GHz (25 mm diameter) to 220 GHz (0.3 mm) has been observed. The  behavior of threshold versus FSR can be understood by considering the dependence of both Q and $A$ versus diameter $D$. Modeling shows that the mode area, $A$, scales approximately like $D^{2/3}$. Over the range of diameters for which the resonator is surface-loss-limited, the Q factor scales approximately linearly with $D$ (i.e., the resonator has an approximately constant finesse over this range). In this surface-loss-limited regime,  Eq.(\ref{eq:thres}) predicts that  the threshold scales like:  $P_{th} \sim D^{-1/3}$. Ultimately, at large diameters, the Q factor will saturate  to a high, constant value. In this limit, Eq.(\ref{eq:thres}) gives: $P_{th} \sim D^{5/3}$. These two regimes are illustrated in Fig.~\ref{fig:schematic} by the trend curves: $D^{-1/3}$ (dashed curve) and  $D^{5/3}$ (dotted curve). The threshold is below 5 mW over a range of FSRs spanning 4.4 GHz to 220 GHz, demonstrating the beneficial effect of surface-loss-limited Q scaling. As an aside, the higher value of threshold for the 2.6 GHz device is a result of this device having a poorer Q factor of around 110 million due to  accidental contamination.

While low threshold power is important, the total required optical pump power for comb operation also depends on threshold. Using   the dimensionless,  coupled, nonlinear equations-of-motion for microcombs in dimensionless form  \cite{Herr}, it can be shown that the power in a comb tooth with mode index $\mu$ is given by:
\begin{equation} \label{eq:combpower}
P_{\mu} = 4 \eta^2  P_{th} \frac{\omega_{\mu} }{\omega_o} \left|a_{\mu} \left( \sqrt{\frac{P}{P_{th}}},  \zeta_{\mu} \right) \right|^2
\end{equation} 
where $\eta$ is the coupling parameter ($\kappa_e/\kappa$), $a_{\mu}$ is the dimensionless field amplitude found by solution of the coupled equations, $P$ is the pump power, $\zeta_{\mu}$ is the dimensionless dispersion parameter,  and $\omega_{\mu}$ ($\omega_o$) is the frequency of the mode $\mu$ (pump).  Here,  any variation in the coupling parameter with $\mu$ is ignored. For a given dimensionless solution set ,$\{a_\mu\}$, threshold power therefore sets the scale for comb power per tooth. Also, lower threshold power improves the pumping efficacy. We would expect these features of threshold to exacerbate the power requirement of comb operation in the microwave regime.

\begin{figure}[t]
\begin{center}
\includegraphics[width=0.45\textwidth]{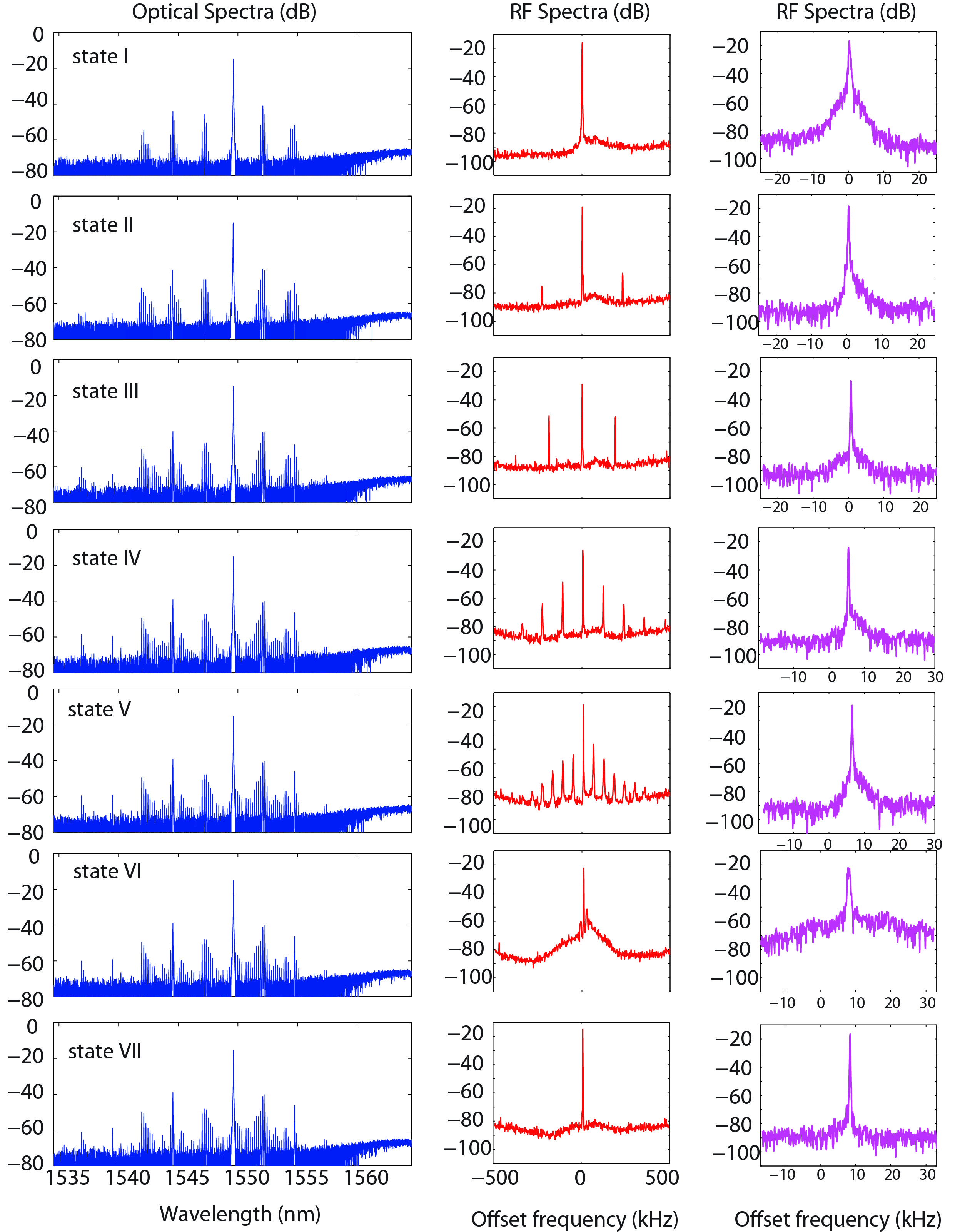}
\caption{\label{fig:phaselock} \textbf{Phase lock dynamics.} Recorded optical spectra (left column) and  RF  spectra (middle and right columns) for different amounts of pump-laser detuning. Detuning is gradually decreased from state I to VII and simultaneously increases the coupled pumping power. The merging beat note spectrum evident in each state of the microcomb is a result of tuning underlying subcomb offset frequencies into coincidence. In State VII there is a sudden collapse of the spectrum into a single, narrow beat note.  The frequency scale for the RF spectra is set with the zero at 21.953 GHz. The span and resolution setting are 1 MHz and 100 Hz for the middle column spectra, and are 50 kHz and 100 Hz for the right column spectra.    }
\end{center}
\end{figure}

As recently described in \cite{Herr}, lower repetition rate combs are prone to unstable microwave beats. This happens because the comb oscillation initiates on a non-native line spacing and grows into a set of secondary combs, each that can feature a distinct offset frequency. In the present system, the ability to fabricate a wide range of FSRs has been used to search for cases in which the resulting dispersion in offset frequencies is low enough to allow pump-detuning-alignment of the underlying sub-comb offset frequencies. We have observed alignment and measured for the first time a locking of the phase across a broad span of comb lines. The effect now described has been observed for a broad range of repetition rates spanning 4 GHz to 26 GHz.

Fig. \ref{fig:phaselock}  shows the comb generation and evolution in a 21.953 GHz comb  as the pump detuning is gradually decreased. Here the pump laser is a CW fiber laser with effective linewidth $\sim$ 1kHz. The optical spectrum  and corresponding RF spectrum (low and higher resolution) are presented at each detuning. The launched pump power in the fiber is 11 mW when the laser is off resonance.  For state I, the coupled pump power is 2.8 mW and  both the primary combs (2.5 nm spacing) and secondary combs (subcombs with native spacing 22 GHz) are visible in the optical spectrum. In this state, only one RF tone is observed. From state II to state V,  more power is coupled into the cavity and the subcombs  grow and begin to overlap, resulting, as described in [18], in additional beat notes. Also, we observe  that the beats tend to merge towards the native-line beat note with continued tuning of the pump towards the optical resonance.   For state VI, the multiple beats merge to create a broad pedestal region around the central beat note. For state VII, the broad pedestal snaps into a single sharp tone with very narrow linewidth and low phase noise.

\begin{figure}[t]
\begin{center}
\includegraphics[width=0.45\textwidth]{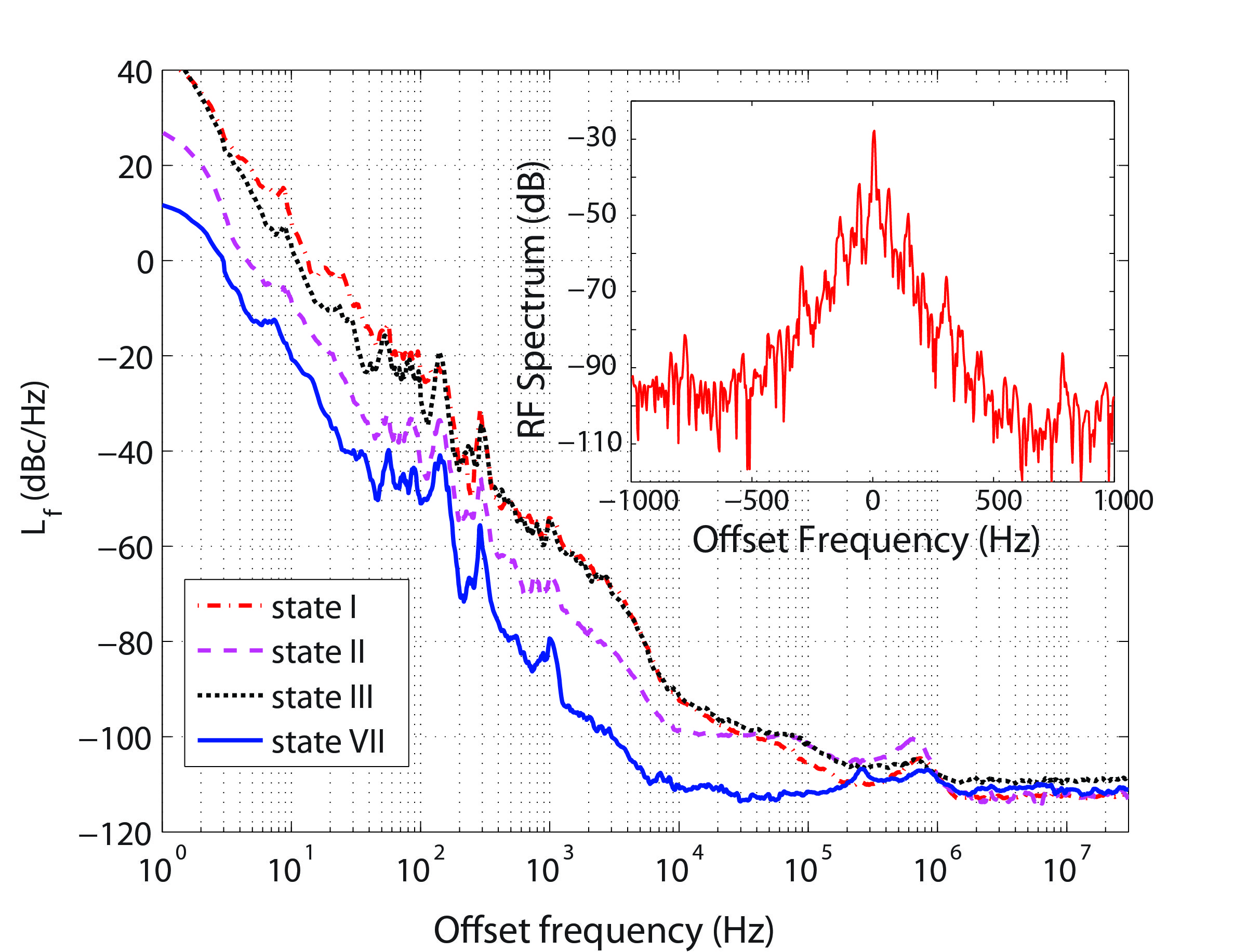}
\caption{\label{fig:PN} \textbf{Phase noise and RF spectrum.} Main panel: Single-side band phase noise of the comb beat note (carrier at 21.953 GHz) for four of the states presented in Fig. \ref{fig:phaselock}. The phase lock state (state VII) shows a very low phase noise of -113 dBc/Hz at 10 kHz offset frequency. Inset:  Spectrum of the RF beat note when the microcomb is in the phase lock state (state VII). Span is set to 2000 Hz and the resolution bandwidth is 10 Hz.  }
\end{center}
\end{figure}

In light of the discussion in  \cite{Herr} we interpret these results as controlled tuning of the underlying subcomb offset frequencies into a state (State VII) of degeneracy. Moreover, the ``snapping" effect and ability to maintain this state over long periods of time suggest that some kind of locking phenomena is at work that self-injection-locks the subcombs into degeneracy once a certain locking range is attained through pump tuning. As further evidence of the phase locking, Fig. \ref{fig:PN} shows a sequence of phase noise spectra taken for the states given in Fig. \ref{fig:phaselock}.  For the phase lock state VII, the phase noise is about 20 - 30 dB lower than the non-phase-locked states. There are approximately 80 comb lines that are phase locked. It is worth noting the very low phase noise level (-113 dBc/Hz at 10 kHz offset for 22 GHz carrier)  is not only a record low phase noise state for a microcomb, but also a level in line with high performance microwave oscillators when operated open loop.

As an aside, certain FSR values are resonant with stimulated Brillouin scattering. For example, a resonator diameter of approximately 6 mm creates a situation in which the FSR is closely matched to the Brillouin shift of approximately 10.8 GHz and this has been used to create high coherence Brillouin lasers \cite{DiskNatureP,SBSOPEX}. The presence of simultaneous Brillouin and parametric oscillation has been reported in \cite{Scott1} and such concomitant oscillation is also observed in the devices of this work.

In summary, microcomb operation over user-determined rates spanning nearly a 100X range was achieved, and, on account of surface-loss-limited Q scaling, the threshold power remained low over this range. Moreover, by analysis of the dimensionless nonlinear dynamical equations for microcombs, we have studied how low threshold performance impacts comb pumping and power scaling.  The present system, having attained 1/2 octave span at 33 GHz repetition rate with 200 mW pump power, is a promising candidate to attain octave span at low repetition rates.  Also, the 2.6 GHz repetition rate is the smallest demonstrated to date for any microcomb. Current efforts are directed towards optimization of dispersion in this system through adjustment of the device structure (thickness and angle control as described in \cite{DiskNatureP}). The ability to study a wide range of FSR values has also enabled operation of the microcombs in a phase-locked state having a record low phase noise for a microcomb.

Beyond the necessity of microwave operation for comb self referencing, ready access to a wide range of microwave repetition 
rates is important for the application of microcombs in a variety of fields including: optical clocks \cite{clock}, astro-physical spectral calibration \cite{Diddams1}, and line-by-line pulse shaping \cite{Purdue}. Moreover, the resonator FSR and hence
repetition rate can be precisely controlled to 1:20000 [18]. Finally, the ability to precisely define these devices on a silicon chip also opens the possibility of integration with waveguides and other devices, and we are actively pursuing this direction.

\begin{acknowledgments}
The authors thank Scott Diddams and Scott Papp for helpful discussions, and also are grateful for financial support under the DARPA QuASAR program. Also, the authors thank the Kavli Nanoscience Institute (KNI)
\end{acknowledgments}


\end{document}